\documentstyle[fleqn,psfig]{tp}

\def\etal{{\it et\thinspace al.\ }}

\def\eg{{e.g}}

\def\lss{{large--scale structure}}

\def\LSS{{Large--Scale Structure}}

\newbox\grsign \setbox\grsign=\hbox{$>$} \newdimen\grdimen \grdimen=\ht\grsign
\newbox\simlessbox \newbox\simgreatbox
\setbox\simgreatbox=\hbox{\raise.5ex\hbox{$>$}\llap
     {\lower.5ex\hbox{$\sim$}}}\ht1=\grdimen\dp1=0pt
\setbox\simlessbox=\hbox{\raise.5ex\hbox{$<$}\llap
     {\lower.5ex\hbox{$\sim$}}}\ht2=\grdimen\dp2=0pt
\def\simgreat{\mathrel{\copy\simgreatbox}}
\def\simless{\mathrel{\copy\simlessbox}}

\begin{document}

\twocolumn[
\title{Clusters as Tracers of Large--Scale\\ Structure} 
\author{Marc Postman,
{\it Space Telescope Science Institute}\\
{\it Baltimore, Maryland 21218 U.S.A.}}
\vspace*{16pt}   

ABSTRACT.\
By virtue of their high galaxy space densities and their 
large spatial separations, clusters are efficient and accurate
tracers of the large--scale density and velocity fields. Substantial
progress has been made over the past decade in 
the construction of homogeneous, objectively derived cluster catalogs
and in characterizing the spatial distribution of clusters. 
Consequently, the constraints
on viable models for the growth of structure have been refined.
A review of the status of cluster--based observations of \lss\
is presented here, including discussions of the second and higher
order moments, the dependence of clustering on richness (mass),
recent and new measurements of bulk flows, and a new constraint 
on the cluster mass function in the range $0.7 \simless z \simless 1$.
\endabstract]

\markboth{Marc Postman}{Clusters as Tracers of \LSS}

\small

\section{Introduction}

The study of the evolution of \lss\ is fundamental
to cosmology. When observations of the galaxy and cluster spatial distributions
are combined with complementary spectroscopic and morphological information,
one can probe the abundance and form of dark matter,
the mean baryon and matter densities, 
the turnover scale in the perturbation power spectrum, 
and the formation epoch of galaxies and clusters.
Clusters of galaxies are particularly well suited
to these studies and, over the past decade, there have been substantial 
breakthroughs in and challenging questions raised about our understanding of 
\lss\ from redshift and peculiar velocity surveys of clusters. 
In this review, I summarize the current constraints on 
the two--point cluster--cluster correlation function (and its dependence 
on cluster richness), the cluster power spectrum, the very large--scale
distribution of clusters, bulk flow measurements, and the evolution
of the cluster mass function. I also emphasize the importance of
characterizing how ones definition of a cluster ultimately affects
the interpretation of observational and N-body simulation results.
I conclude by highlighting several exciting cluster research
programs now (or soon to be) underway which will provide new levels of accuracy 
in defining the relationship between the large--scale distribution 
and properties of clusters and the underlying astrophysics responsible for   
their formation and evolution.

\section{The Case for Clusters As Tracers of LSS}

There are several powerful arguments in support of using clusters to
trace \lss\ in the universe. 
Clusters lie $10 - 30$ times farther apart
than galaxies, with intercluster separations in the range
$30 - 100/h$ Mpc. This fact immediately makes them promising candidates
for tracing \lss\ because a relatively small sample can be used
to probe great distances. 
Clusters are probably ``close" to their initial formation positions 
($\pm10 (v_{pec}/1000 \ {\rm km/s}) (t_o/10 \ {\rm Gyr})\  h^{-1}$ Mpc) 
because the typical peculiar velocity of an individual cluster is
500 km/s or less (Watkins 1997). Furthermore, clustering of clusters is
significantly amplified with respect to that for galaxies. This is, in part,
due to the fact that clusters are located at highest peaks in matter 
distribution ($\rho_{CL}/\rho_c \simgreat 100$). This enables one to study
features of the distribution that might otherwise be too weak to
measure from galaxy redshift surveys (\eg, correlations on scales
in excess of $30h^{-1}$ Mpc) and features which are dependent on the
higher moments of the galaxy distribution. High order moments, for example,
are highly sensitive to the details of most biasing prescriptions. 
In addition, the dark matter on typical intercluster distance 
scales is likely still in the quasi-linear 
(or only mildly non-linear) regime and, thus, it is easier to associate present
epoch constraints on the shape of the perturbation spectrum with
its primordial counterpart. N-body simulations 
show that clusters are unbiased tracers of the underlying velocity field 
(Strauss \etal\ (1995); Gramann \etal (1995)) 
and provides further theoretical justification
for the extensive observational efforts to measure 
large--scale flows (see Section~\ref{vpec}).  
The observations have been motivated by the fact that
relative redshift-independent distances to clusters can be
measured to accuracies $\simless$5\% owing to the large number of galaxies
in each system. Clusters are also the exclusive sites for application of 
a number of secondary distance indicators
(\eg, BCGs, SZ Effect) and large elliptical populations 
(allowing accurate fundamental plane measurements). 
Lastly, clusters can be easily detected out to $z \sim 1$ 
(and beyond, if NIR data are available) owing to their high
density contrast, red elliptical population, and bright member galaxies. 
This enables important constraints to be made on 
the evolution of \lss\ and $\Omega_o$.

\section{Identifying Clusters}

There are three basic ways to ``define" what one means by a cluster
of galaxies. They are:

\underbar{The Physical Definition} (at $z \sim 0$): Gravitationally bound, 
virialized system of dark matter, gas, and galaxies with a mass of at 
least $\sim10^{14}$ M$_{\odot}$ inside a region $\sim1/h$ Mpc in radius.

\underbar{The N-body Definition}: Peaks in a dark-matter dominated
density field, perhaps
satisfying additional constraints such as a minimum velocity dispersion,
$\sigma$, and/or minimum value of $\sigma \rho$. The peaks can be located
via Friends-of-Friends (FoF) or Local Overdensity (LO; nearest $n$ particles) 
algorithms. The locations in real space are known precisely but the precise 
relationship between the dark and luminous matter distributions is poorly
understood at present.

\underbar{The Observational Definition(s)}: These can vary substantially 
depending on number of available positional coordinates per galaxy (2D/3D) 
and on the wavelength of the 
survey (opt, IR, x-ray). Modern optical/IR searches 
identify density enhancements using FoF, LO, or matched filter 
algorithms in either 2 or 3 dimensions. X-ray searches look for sources
with extended emission and/or cross-correlate all x-ray sources with
optically-selected galaxy catalogs. Complications include projection 
effects (2D), redshift distortions (3D), sensitivity variations across the 
survey, and for the Abell (1958), Abell, Corwin, Olowin (ACO; 1989), and 
Zwicky \etal (1968) catalogs, some degree of human error.

Because the clustering properties of clusters will depend on how they were
selected (\eg, Eke \etal 1996), 
any comparisons between observational datasets or
between observations and simulations must quantify
and correct for biases introduced by the selection process.
This task is tractable for samples derived using well-understood
and quantifiable selection criteria.

\subsection{Cluster Catalogs}

There have been at least 14 original cluster catalogs constructed to date and 
Table~\ref{tab:1} summarizes their basic parameters.
The impact of high speed, memory-rich computers and digital data is readily 
apparent: subsequent to 1992, cluster detection relies exclusively on 
the objective
application of algorithms with accurately quantifiable selection functions
(``Bulleted" catalogs). None the less, results from the visually derived
Abell and ACO catalogs are still widely cited because they have been the only
all sky cluster surveys available to date. The Zwicky cluster catalog, by
contrast, is only infrequently used largely because their cluster finding 
procedure, visual identification of global isodensity contours twice as
high as the mean contour, is fraught with many pitfalls
including enhanced sensitivity to plate-to-plate photometric 
zeropoint variations. 
The cluster identification prescription developed by Abell is a bit more
robust and, indeed, the automated APM catalog is based upon a modified 
version of this approach.

\begin{table*}
\caption[]{Catalogs of Clusters of Galaxies}
\centering
\begin{tabular}{rl|c|c|r|r}
\hline \hline
 &         & Detection & Approximate    & Number of & Number with \\
 & Catalog & Passband  & Redshift Range & Clusters  & Spec. Redshifts \\
\hline
 & Abell (1958)                  & Opt. & $z \simless 0.3$ & $\sim 2700$ & $\sim 1300$\\
 & Zwicky \etal (1968)           & Opt. & $z \simless 0.3$ & $\sim 9000$ & $\dots$ \\
$\bullet$& Shectman (1985)       & Opt.  & $z \simless 0.3$ & $\sim 650$  & $\dots$ \\
 & Gunn, Hoessel, Oke (1986)     & Opt./NIR & $z \simless 1$   & $\sim 400$  & $\sim 50+$ \\
 & Abell, Corwin, Olowin (1989)  & Opt. & $z \simless 0.3$ & $\sim 1350$ & $\sim 250+$ \\
 & Couch \etal (1991)            & Opt. & $z \simless 0.6$ & $\sim 100$  & $\sim 20$ \\
$\bullet$& Henry \etal (1992) EMSS & X-Ray & $z \simless 0.6$ & $\sim 95$ & $\sim70+$ \\
$\bullet$& Lumsden \etal (1992) EDCC & Opt. & $z \simless 0.3$ & $\sim 700$  & $\sim 100+$ \\
$\bullet$& Dalton \etal (1994a) APM  & Opt. & $z \simless 0.3$ & $\sim 1000$ & $\sim 360+$ \\
$\bullet$& Postman \etal (1996) PDCS & Opt./NIR & $z \simless 1$   & $\sim 80$   & $\sim 20+$ \\
$\bullet$& Scodeggio, Olsen, \etal (1998) EIS & Opt./NIR & $z \simless 1$   & $\sim 250$   & $\dots$ \\
$\bullet$& Rosati \etal (1998) RDCS & X-Ray & $z \simless 0.8$ & $\sim 70$ & $\sim 60+$ \\
$\bullet$& Vikhlinin \etal (1998) & X-Ray & $z \simless 0.6$ & $\sim 200$ & $\dots$ \\
$\bullet$& Boehringer \etal (1998) REFLEX & X-Ray & $z \simless 0.3$ & $\sim 450$ & $\sim 380+$ \\
\hline
\end{tabular}
\noindent $\bullet$ = Automated Catalog
\label{tab:1}
\end{table*}

Postman \etal (1986) demonstrated that the angular correlation
functions of Abell and Zwicky clusters agree when appropriate sub-samples 
of each catalog are chosen
(corresponding to the spatial regime where the different algorithms
identify similar types of clusters).
Without such careful comparison, the $\omega_{cc}(\theta)$ from Abell and
Zwicky clusters differ significantly. A similar level of discrepancy has
existed between the spatial correlation functions for the Abell/ACO clusters
and that for the APM clusters (\eg, Dalton \etal 1992; Postman \etal 1992). 
These differences are, in large part,
due to differences in the respective selection criteria which result
in different minimum richness limits in the catalogs (Bahcall \& West 1992). 
To a lesser degree, differences in the \lss\ statistics for
the Abell/ACO and APM catalogs are due to projection effects which appear to
be strongest in the original Abell catalog (Sutherland 1988; Soltan 1988;
Efstathiou \etal 1992).
A cross-correlation of the ACO and APM catalogs in a sector of sky
covered by both reveals that about 75\% of the
ACO clusters with $m_{10} \le 16.5$ (corresponding to $z \simless 0.1$)
have an APM counterpart. The percentage drops to 50\% for $m_{10} < 18$.
Conversely, about 70\% of the APM clusters with $m_X \le 16.9$ have
ACO counterparts; 50\% match when $m_X \le 17.8$. Some of the failed matches
are due to differences in the respective depths of the two catalogs and most
of the remainder are due to differences in richness cuts imposed by
the different detection schemes. 

\begin{figure*}
\centering\mbox{\psfig{figure=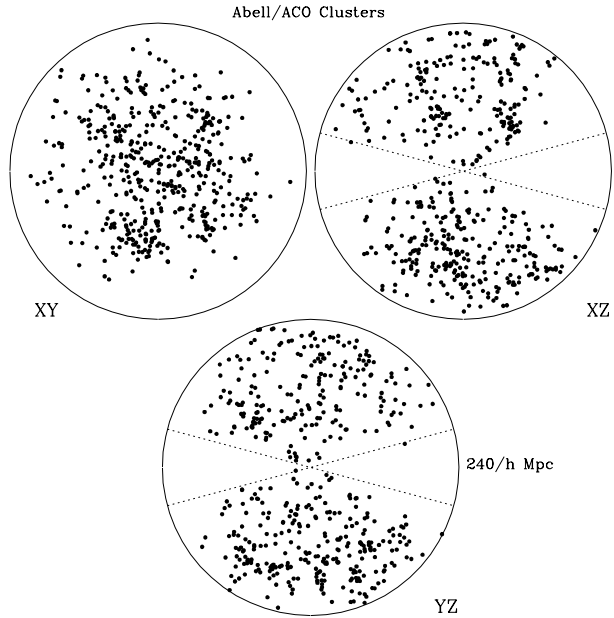,height=3.7in}}
\caption[]{The spatial distribution of a nearly volume limited sample
of $\sim 480$ Abell and ACO clusters (RC $\ge$ 0). The axes are aligned with
the Galactic coordinate system. The zone of avoidance at 
$|{\rm b}| = 13^{\circ}$ is shown.}
\label{xyzplot}
\end{figure*}

The spatial distribution of a nearly volume limited ($z \le 0.08$), 
all--sky sample of $\sim480$ Abell/ACO clusters is shown in 
Figure~\ref{xyzplot}. The space density of southern ACO clusters is
about a factor of 2 larger than that of northern Abell clusters primarily
due to the more sensitive IIIaJ emulsion used in the southern survey and 
to the presence of the Shapley supercluster ($z \approx 0.047$) in the south. 
The ACO cluster space density is quite similar to that found in the APM 
cluster catalog, about $2.4 \times 10^{-5}h^3$ Mpc$^{-3}$. 

\section{Second Order Cluster Correlations}

Second order statistics such as the two--point
spatial correlation function, $\xi(r)$, 
and its Fourier transform, $P(k)$, have been widely used to constrain
the clustering properties of clusters. Although competing structure
formation models can occasionally yield quite similar predictions for
$\xi(r)$ and $P(k)$, they are none the less robust measures of clustering
strength which provide basic information about the underlying matter 
distribution. They are also computationally straight forward to compute
and substantial work has been dedicated to identifying the optimal
estimators for these functions (Landy \& Szalay 1993; Hamilton 1993; 
Peacock \& Dodds 1994; Landy \etal 1996, Tegmark \etal 1998).

\subsection{The Cluster Auto--Correlation Function}

\begin{figure*}
\centering\mbox{\psfig{figure=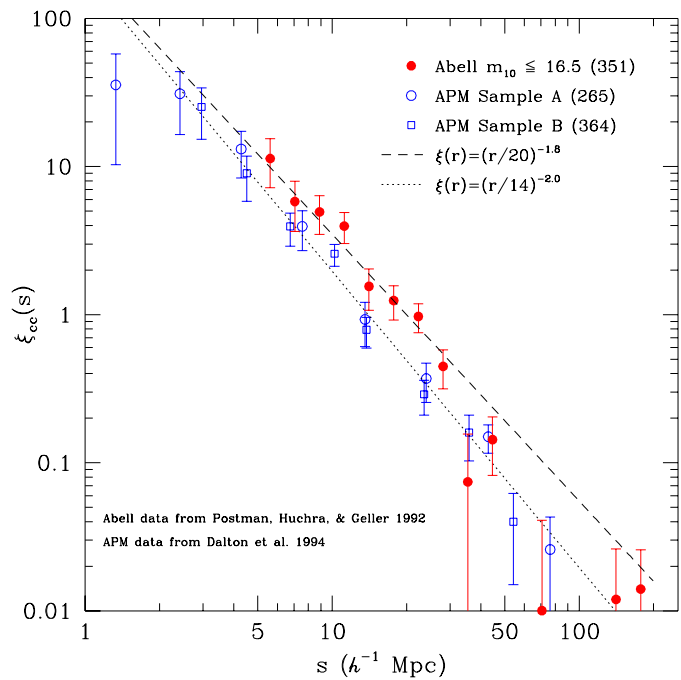,height=3.7in}}
\caption[]{The spatial correlation function for APM and Abell clusters,
respectively. The best-fit power laws are shown.}
\label{xicc}
\end{figure*}

The cluster--cluster 
spatial correlation function is often fit to a power-law of the form
$$\xi_{cc}(r) = \left( r \over r_o \right)^{-\gamma}$$ 
This appears to be a reasonable model over $5 \le r \le 35h^{-1}$ Mpc
and results from several independent 
cluster catalogs yield $1.8 \le \gamma \le 2.2$. 
Determinations of $\xi_{cc}(r)$ from the Abell and APM catalogs are shown in
Figure~\ref{xicc}. Although the amplitude of $\xi_{cc}(r)$ differs by factors
between 6 to 30 from that for the galaxy autocorrelation function
(with the precise ratio dependent on the sample compositions), the shape of
the cluster and galaxy correlation functions are very similar. This similarity
in shape is also seen in the respective power spectra, at least on scales
less than 70$h^{-1}$ Mpc where the signal-to-noise ratio is high.

The most stringent constraints that $\xi_{cc}(r)$ can place
on structure formation models come from addressing the following questions:
\begin{quote}
\noindent Do $r_o$ and $\gamma$ change with the mass-scale of 
the systems being studied?

\noindent Are there deviations from a power law (\eg, $\xi_{cc}(r) \le 0$)
and on what scales?

\noindent How do the amplitude and shape of $\xi_{cc}(r)$ vary 
with cosmic time?
\end{quote}
Expectations are that, to some degree, $r_o$ must depend on the mass
of the cluster: if structure grows by gravitational amplification
of fluctuations in a Gaussian field, then collapsed objects form near
peaks in this field and their clustering will depend on the height of
the peak (Kaiser 1984; Barnes \etal 1985). 
Early data on $\xi_{gg}(r)$ and $\xi_{cc}(r)$ led
Szalay \& Schramm (1985) to propose a ``scale-invariant" form 
given by
$$\xi_i(r) = {1 \over 3} (r/d_i)^{-1.8} $$
and thus $r_o = 0.54 d_i$
where $d_i$ is the mean interobject separation. 
Szapudi, Szalay, \& Bosch\'an (1992) demonstrated that amplification 
is a consequence of enhanced weighting of dense regions when deriving
the higher moments of the density field (\eg, $\xi_{cc}(r)$).
It is important to emphasize that the existence of a scale dependence
to the correlation length does not imply that galaxies and clusters cannot
both be tracers of the distribution of \lss. Rather, it suggests that
these two classes of systems trace the underlying matter differently.

Bahcall \& West (1992; BW92) proposed that data for a wide range of catalogs
of clusters and galaxies satisfy a scale-invariant relationship between
$r_o$ and $d_i$ of the form $r_o = 0.4 d_i$. 
Croft \& Efstathiou (1994), however, could not reproduce a relationship between
$r_o$ and $d_c$ (intercluster separation) as strong as the BW92 result
using SCDM N-body simulations. They found a trend which showed little
dependence of $r_o$ on $d_c$ when $d_c \simgreat 30h^{-1}$ Mpc.
A subsequent analysis by Croft \etal (1997), using an extended subset of
the richest APM clusters, suggests that these systems show only a weak
dependence of $r_o$ on $d_c$ that is consistent with low density
CDM models. In contrast,
Walter \& Klypin (1996) were able to reproduce a relationship between
$r_o$ and $d_c$ which is as strong as the 
BW92 result from CHDM N-body simulations. However, those same simulations
predict a dramatic decrease in the comoving space density of clusters
as one looks back to $z = 0.5$, a prediction which is clearly not
consistent with present observations (Postman \etal 1996;
Carlberg \etal 1997). See also the results
based on the Virgo Consortium simulations (Colberg \etal, these
proceedings).

\begin{figure*}
\centering\mbox{\psfig{figure=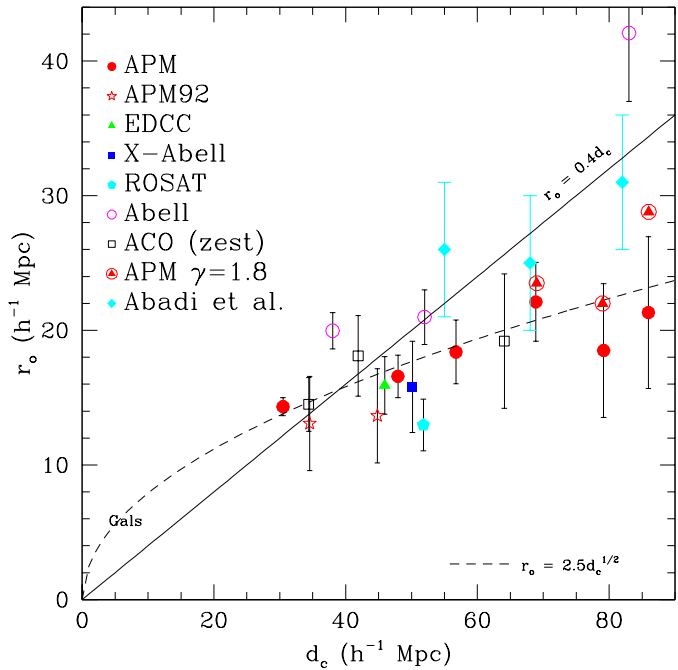,height=3.7in}}
\caption[]{The spatial correlation length as a function of the intercluster
separation for different cluster samples. At large $d_c$, we show
the APM results when the slope, $\gamma$, is constrained to be 1.8.
The relationships $r_o = 0.4d_c$ and $r_o = 2.5\sqrt{d_c}$ are shown for 
comparison. Based on results from Bahcall \& West (1992),
Croft \etal (1997), Abadi, Lambas, \& Muriel (1998), and this review.}
\label{rodc}
\end{figure*}
 
Figure~\ref{rodc} summarizes the current situation. Part of the
scatter in the figure is due to inconsistent comparisons between
authors. The BW92 dependence
of $r_o = 0.4d_c$ was based on power law fits to $\xi_{cc}(r)$ with
$\gamma$ constrained to be 1.8. For 
the APM results, Croft \etal (1997) allow $\gamma$ to
be a free parameter and often obtain fits with $\gamma \simgreat 2.2$
for $d_c > 60h^{-1}$ Mpc. Since there exists a significant covariance
between $r_o$ and $\gamma$, such comparisons must really be done
at a common slope value. Indeed, when one fixes $\gamma$ at 1.8,
the APM $r_o$ at $d_c > 60h^{-1}$ Mpc do increase as shown.
A more substantial cause for scatter in Figure~\ref{rodc} is
demonstrated by Eke \etal (1996) who find that $r_o$ can
vary by up to 50\% depending on the precise cluster identification procedure.
They could reproduce either a strong or weak dependence
of $r_o$ on $d_c$ depending on which cluster identification method was used
and are able to reconcile the BW92 and the Croft \etal (1997) findings
as consequences of the different selection procedures used by ACO
and by the APM team.

In sum, a weak dependence of correlation amplitude on intercluster
separation (and richness) is fairly well-established ($r_o \propto \sqrt{d_c}$
or $r_o \approx 0.2d_c$) both theoretically and observationally. 
The observational evidence for a stronger dependence (\eg, $r_o = 0.4d_c$) 
is from the angular clustering properties of Abell RC $\ge 2$ clusters,
the clustering of x-ray bright Abell clusters (Abadi, Lambas, \& Muriel 1998)
and the supercluster correlation function (Bahcall \& Burgett 1986) ---
all of which are derived from the Abell and ACO catalogs.
The APM cluster results for $d_c > 60h^{-1}$ Mpc 
are based on $<60$ clusters and, hence, are subject
to possible systematic effects (as is any small catalog). The results
for $d_c > 60$ thus require confirmation from larger redshift surveys
({\it e.g.}, the extended APM, 2dF, and Sloan Digital Sky Survey [SDSS]). 
In any event,
careful attention needs to be paid by both observers and theorists
to the not so subtle effects of the cluster selection process on
\lss\ statistics before any physical inferences are made based on 
the observed relationship between $r_o$ and $d_c$.
 
\subsection{The Zero Crossing of $\xi(r)$}

Deviations in $\xi(r)$ from a single power law behavior 
are, in principle, sensitive tests for the shape of the
primordial fluctuation power spectrum. One such deviation is the
scale at which the correlation amplitude goes to zero 
(Klypin \& Rhee 1994; KR94).
The amplitude of $\xi_{cc}(r)$ appears to be positive at least
out to 40$h^{-1}$ Mpc and possibly out to scales of 60$h^{-1}$ Mpc
(Postman \etal 1992; Olivier \etal 1993; KR94;
Dalton \etal 1994b; Boehringer \etal 1998). However, on scales
from $60 - 100h^{-1}$ Mpc, it is very unlikely ($2-3\sigma$ level)
that $\xi_{cc}(r) > 0$.
This result is seen in the Abell/ACO, APM, and REFLEX (x-ray
selected) cluster catalogs. Systematic effects, such as the integral
constraint for a finite sample (which forces $\xi(r)$ to 
eventually become negative) or small errors in the mean cluster
number density, appear not to be large enough to fully explain
this zero crossing. {\it One important implication of this result may be that 
$\xi_{gg}(r)$ is also positive out to at least 40$h^{-1}$ Mpc}
(Szapudi \etal 1992). Indeed, preliminary results from the 2dF redshift
survey (Maddox, these proceedings) 
and the ESO Slice Project (Guzzo, these proceedings) both find
positive $\xi_{gg}(r)$ out to at least 35$h^{-1}$ Mpc. Such observations
put severe constraints on CDM models. As noted by KR94,
$\Lambda$CDM models (\eg, Kofman, Gnedin, \& Bahcall 1993)
predict a zero crossing at $r_z = 16.5(\Omega h^2)^{-1}$ Mpc. 
If the above observations hold up, then this suggests
that $\Omega h$ lies in the range $0.28 - 0.41$.
 
\subsection{The Cluster Power Spectrum}

\begin{figure*}
\centering\mbox{\psfig{figure=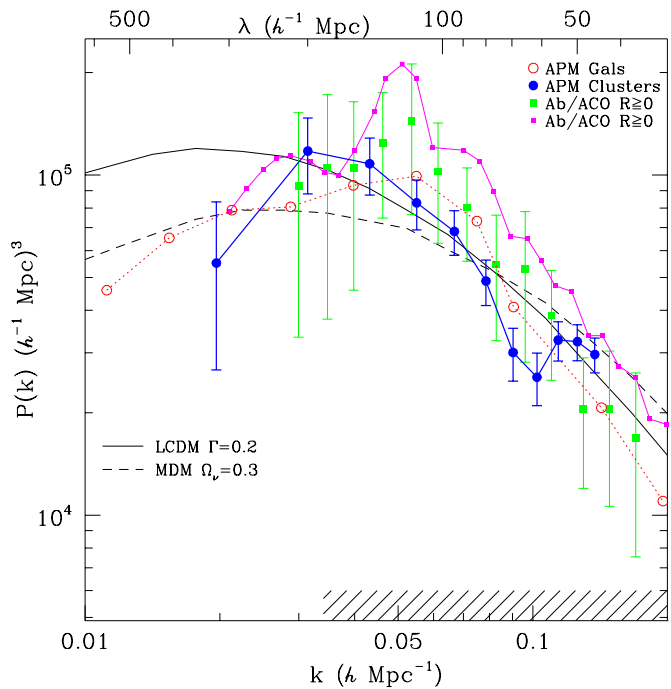,height=3.7in}}
\caption[]{Power spectra for APM galaxies (Maddox \etal (1996)), 
APM clusters (Tadros \etal (1998)), and Abell/ACO
clusters (large squares are Retzlaff \etal (1997); small squares
are Einasto \etal (1997)). The galaxy power spectrum has been normalized 
to match the amplitude of the APM cluster power spectrum. $P(k)$ for 
two models also shown.}
\label{pkplot}
\end{figure*}

The power spectrum of clusters, $P(k)$, provides a complementary constraint
on their clustering properties: broad features in the correlation function
are narrow in Fourier space and vice versa. Furthermore, errors in $P(k)$
are easier to estimate correctly
and the results are somewhat less sensitive to uncertainties in the mean
space density of clusters than those for $\xi(r)$. 
Current constraints on the cluster power spectrum are
shown in Figure~\ref{pkplot}. The shape of cluster power spectrum, like
it's inverse Fourier transform $\xi(r)$, is consistent with the shape
of the power spectrum of optical, IRAS, and radio galaxies at
$k > 0.04h$ Mpc$^{-1}$ (Peacock \& Dodds 1994; Einasto \etal 1997;
Retzlaff \etal 1997; Tadros \etal 1998). This suggests, again, that clusters
and galaxies are tracing similar perturbations in the matter distribution. 
The turnover in $P(k)$ is detected (but not with high significance) for
$k < 0.03h$ Mpc$^{-1}$. The SDSS, 2dF, and other large redshift surveys 
should eventually yield dramatically improved constraints on the turnover,
a feature which depends upon the horizon scale at the epoch of matter-radiation
equilibrium.

The amplitude of $P(k)$ for Abell/ACO clusters is, on average, 
a factor of $\sim 2 - 3$ higher than that for APM clusters, consistent
with differences seen in their respective $\xi(r)$.
In turn, the APM cluster $P(k)$ amplitude is about $6 - 8\times$
higher than that derived for galaxies in the Las Campanas Redshift
Survey (Lin \etal 1996; LCRS). The observed shape of $P(k)$ is reasonably
well represented by MDM models
($0.2 \simless \Omega_{\nu} \simless 0.3$),
low--density CDM models ($\Omega h \sim 0.3\pm0.1$), and/or
$\Lambda$CDM models ($\Lambda \sim 0.3$) (Borgani \etal 1996).
The apparently strong feature seen at $120 \pm 15h^{-1}$ Mpc in the
Abell/ACO $P(k)$ (Einasto \etal 1997)
is {\it not} seen in the $P(k)$ derived from either APM clusters
(Tadros \etal 1998) or REFLEX x-ray selected clusters (Boehringer \etal 1998). 
A subsequent analysis of the Abell/ACO $P(k)$ by 
Retzlaff \etal (1998) find that the feature is not
statistically significant when sample variance is properly accounted for.
None the less, a statistically significant feature is detected in the 2D 
LCRS $P(k)$ at around $100 h^{-1}$ Mpc (Landy \etal 1996) and spikes 
continue to be found in the galaxy redshift distribution on
similar scales in new pencil beam surveys (Broadhurst \etal 1995).
These features, which are not reproduced by most non-baryonic 
matter dominated models, are presumably due
to characteristic scales of voids and sheets in the galaxy distribution.
The lack of a significant detection of this feature in cluster power
spectra may be a consequence their sparser sampling of the density field.
For instance, a direct comparison between the galaxy distribution from the 
extended (R $\le\ 15.4$, $N_{gal} \sim 6000$) CFA redshift survey 
(Geller 1998) and the Abell cluster distribution
in the same volume, reveals that several prominent features in the galaxy
distribution which contribute to the peaks originally found by Broadhurst
\etal (1990) are not traced by the clusters. 

Narrow, large amplitude features in $P(k)$ are surprising yet intriguing. 
There is presently no theoretical consensus on the origin of preferred 
scale lengths.  Eisenstein \etal (1998) hypothesize that such excess power could
be a consequence of baryonic acoustic oscillations in adiabatic models.
However, they note that this would require a substantial error
in currently favored values of cosmological parameters.
Einasto \etal (1997) simply conclude that our present understanding of
the formation of \lss\ requires substantial revision.

\section{High-Order ($N \ge 3$) Statistics}

Higher order statistics potentially provide some of the best constraints
on the degree of biasing (\eg, Jing 1997) and, thus, on the reliability of 
clusters as tracers of the mass. The high-order moments of cluster
distributions have already been shown to be non-zero. For example,
ACO clusters exhibit hierarchical clustering behavior given by
\begin{eqnarray}
\xi_N(r_1,...,r_N) = \nonumber \\
\sum_{\alpha}Q_N^{(\alpha)} \sum_{ij} \Pi^{N-1}\xi_2(r_{ij}) \nonumber
\end{eqnarray}
up to 6th order (Cappi \& Maurogordato 1995) with $Q_3 \approx 1.0$. 
APM clusters display a similar hierarchical behavior, at least up to 4th
order (Gaztananga, Croft, \& Dalton 1995).

The deprojected $S_N$'s ($\xi_N = S_N \xi_2^{N-1}$) 
for APM galaxies (Gaztanaga 1994) and ACO clusters are quite
similar in amplitude which suggests that ACO clusters and APM galaxies 
are sampling the same underlying matter distribution but that the
biasing between galaxies and clusters is non-linear.
The $S_{3,4}$ values for ACO and APM clusters
are (3.1, 22) and ($\sim2,\ \sim8$), respectively. 

\section{Very Large Scale ($>200h^{-1}$ Mpc) Structure }

The large intercluster spacing and the enhanced amplitude of their clustering
makes the study of structure on very large scales possible, in principle.
In practice, the signals on these scales are small and errors in 
modeling systematic effects such
as photometric zeropoint variations, sample variance, 
or Galactic reddening can yield artificial signals of comparable amplitudes. 
Tully (1987) first proposed the detection of very
large-scale alignment of the local Abell cluster distribution
with the Supergalactic plane. Postman \etal (1989) countered that this effect
(based on available data at the time) was not statistically significant
($<2\sigma$) and could be expected from sample with the observed $\xi(r)$. 
Tully \etal (1992) extended their analysis to a full-sky sample of Abell
and ACO clusters and still found an alignment with the Supergalactic plane,
claimed to be significant at the 6-sigma level, extending to scales
of 450$h^{-1}$ Mpc. The structure is only a small amplitude
fluctuation ($\delta\rho/\rho \simless 0.015$), if indeed real.
Scaramella (1992) also used Abell/ACO clusters to study power
on 600$h^{-1}$ Mpc scales and found relatively low values for 
for density fluctuations, consistent with limits on the fluctuations
in the cosmic microwave background (CMB).
The study of very large scale structure performed directly from cluster
redshift surveys will not likely advance much further until automated
wide--area, homogeneous cluster catalogs, like those expected from the SDSS
or the extended APM survey ($\sim$900 clusters; Tadros 1998), become available. 

\section{The Large-scale Velocity Field}\label{vpec}

A complementary approach (and perhaps a more promising one given current
cluster catalogs) to studying very \lss\ using clusters
is through the mapping of the large--scale velocity
field.  Currently, at least 7 independent cluster-based peculiar velocity 
surveys, all reaching scales of 100$h^{-1}$ Mpc or larger,
are either complete or in progress (see Table~\ref{tab:2}). 
Inferences about the underlying mass distribution from
peculiar velocity surveys are less susceptible to incompleteness effects
and radial density gradients
than those from redshift surveys. However, peculiar velocity surveys
require highly accurate photometric and spectroscopic calibrations and
extremely homogeneous data (see Strauss, these proceedings). Careful
characterization of the systematic errors and the
effects of sparse sampling are also 
required (\eg, Lauer \& Postman 1994; Feldman \& Watkins 1994).
\begin{figure*}
\centering\mbox{\psfig{figure=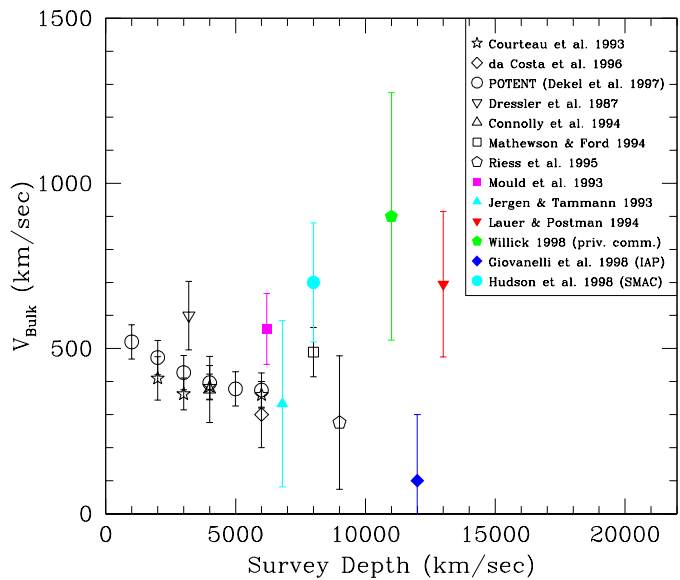,height=3.7in}}
\caption[]{The amplitude of derived bulk flows from recent galaxy and
cluster based peculiar velocity surveys. The results from cluster-based
surveys are indicated by the filled data points.}
\label{vbulk}
\end{figure*}
Figure~\ref{vbulk} summarizes the current constraints on bulk flow amplitudes
from both galaxy and cluster based surveys. The constraints on the
largest scales are nearly all from cluster-based surveys. 
Included in the plot are two new results. The exciting results of a 
700 km/s flow at 8000 km/s depth from 
Hudson \etal are discussed elsewhere in these proceedings. 
Willick (1998) reports the measurement of $v_{Bulk} = 900\pm375$ km/s
(1$\sigma$ error) in the redshift range $9000 \le cz \le 12,000$ km/s
based on a Tully-Fisher survey of 15 rich Abell clusters. 
Neither of these two results are consistent with the direction
of Lauer \& Postman (1994; LP) result. They may, however, be consistent with 
each other. Indeed, no other work to date has corroborated the LP
bulk flow (see also Wegner \etal (1998) and Saglia, these proceedings).
This may suggest that either the original LP BCG sample is a statistical
fluke or an additional parameter is required for accurate BCG distance
estimation (\eg, Hudson \& Ebeling 1997). 
The extended BCG survey by Lauer, Postman, \& Strauss (1999) will
provide a good test. In contrast, Dale \etal (1997)
find no evidence for a large bulk flow at 8000 km/s. The inconsistent
results for the amplitude and direction of large--scale bulk flows argue that
the convergence scale is not yet well--established. However, the quality
and quantity of data from the on-going surveys, including promising
results from space-based SBF studies, should be sufficiently good that
much better constraints will be available within the next two years or so.

\begin{table*}
\caption[]{Current Cluster-based Peculiar Velocity Surveys}
\centering
\begin{tabular}{l|c|r|c|r}
\hline \hline
Investigators & D.I. & $N_{clus}$ & Depth (km/s) & $\sigma_D$/cluster \\
\hline
Dale \etal (1997)            & TF  & $\sim50$  & 18,000 & $\sim7$\% \\
Gibbons, Fruchter, Bothun (1998)&FP&       20  & 11,000 & $\sim6$\% \\
Hudson \etal (1998; SMAC)    & FP  & $\sim60$  & 12,000 & 7\% \\
Lauer \& Postman (1994)      & BCG &      119  & 15,000 & 16\% \\
Lauer, Postman, Strauss (1999)& BCG& $\sim500$ & 24,000 & 16\% \\
Tonry \etal (1998)           & SBF &       11  & 10,000 & $\simless 5$\% \\
Wegner \etal (1998; EFAR)    & FP  &       84  & 15,000 & 8\% \\
Willick (1998)               & TF  &       15  & 12,000 & $\sim5$\% \\
\hline
\end{tabular}
\label{tab:2}
\end{table*}

\section{The Cluster Mass Function at $z > 0.7$}

The advent of the Keck telescope and the Low-Resolution Imaging
Spectrograph have revolutionized spectroscopic surveys of distant
($z > 0.7$) clusters. This capability has now enabled us
(Oke, Postman, \& Lubin 1998) to provide a preliminary constraint
on the normalization of the cluster mass function (CMF)
in range $0.76 < z < 0.92$ based on data for 3
clusters and between 22 and 36 cluster members for each system.
The specifics of the mass estimation techniques are described in 
Postman, Lubin, Oke (1998). The new constraints on the CMF
are based on the clusters CL1324+3011 ($z=0.76$), CL1604+4304 ($z=0.90$), and 
CL1604+4321 ($z=0.92$). The latter two are part of a supercluster.
All 3 clusters are from the Gunn, Hoessel, Oke (1986; GHO) catalog and their 
kinematic masses are all in excess of $4.5\times10^{14}h^{-1}{\rm M}_{\odot}$
within their central $1h^{-1}$ Mpc regions. If we make the quite
conservative assumption that these are the {\it only} 3 clusters
this massive within the entire GHO catalog ($\sim72$ deg$^2$), then
we find that, for $\Omega_{\circ} = 0.2$, a lower limit on the
CMF in the range $0.7 < z < 1$ 
\begin{eqnarray}
N(\ge{\rm M}=4.5\times10^{14}h^{-1}{\rm M}_{\odot}) > \nonumber \\ 
\medskip 1.1\times10^{-7} h^3 {\rm Mpc}^{-3} \nonumber 
\end{eqnarray}
This constraint is consistent with estimates made by 
Bahcall, Fan, \& Cen (1997) and provides additional observational support
for a low-density universe. While this constraint is relatively crude
($\pm$ factors of $2 - 20$), the discovery of similarly massive systems
(\eg, MS1054-03 and MS1137+66) at similarly high redshifts will likely
continue to grow as observations of distant clusters progress.

\section{Cosmological Implications and Future Developments}

The general conclusions one can draw from the ensemble of cluster
data and simulations discussed above are that
\begin{enumerate}
\item Clusters are reliable tracers of the underlying mass but trace it
differently from galaxies. In particular, clusters trace the 
\lss\ sparsely since they are relatively rare objects. 
The biasing between galaxies and clusters is non-linear and is 
dependent on their intrinsic properties (\eg, the central
mass of the cluster).

\item The statistically significant power seen in the cluster
distribution on scales between $30 - 60h^{-1}$ Mpc implies
that galaxies are also likely to exhibit correlations on
the same scales. Indeed, the larger galaxy redshift surveys
(LCRS, 2dF, ESO Slice survey) now confirm that $\xi_{gg}(r) > 0$
at least to $35h^{-1}$ Mpc.

\item The cluster observations seem to favor MDM models 
($0.2 \simless \Omega_{\nu} \simless 0.3$),
low--density CDM models ($\Omega h \sim 0.3\pm0.1$), and/or
$\Lambda$CDM models ($\Lambda \sim 0.3$). The exception would
be if the large--amplitude, large--scale bulk flows persist, in which
case somewhat higher values for $\Omega$ are required.  

\item Massive (few $\times 10^{14}h^{-1}{\rm M}_{\odot}$) clusters
exist at $z > 0.7$ in an abundance that is hard to reconcile
with $\Omega_{\circ} = 1$ models.
\end{enumerate}
There is still much we need to learn about \lss\ formation and evolution and
a number of exciting developments over the next 3 to 5 years will help.
New, larger objective cluster catalogs will soon be available
from surveys such as the SDSS, 2dF, and extended APM.
Using them, we should be able to constrain the cluster power spectrum
with unprecedented accuracy to scales approaching 1 Gpc. These catalogs
will also enable more extensive, direct
comparisons between the galaxy and cluster 
distributions in identical volumes and will allow us to
establish, with significantly better accuracy, the dependence
of the clustering properties of clusters on their intrinsic parameters.
Joint x-ray/optical cluster searches (\eg, Donahue \etal 1999)
should elucidate the nature of cluster evolution at intermediate
redshifts ($z \simless 1$). Deep, wide-area galaxy surveys (\eg,
Postman \etal 1998b; Jannuzzi, Dey, \etal 1998) will provide important
and new measurements of the evolution of \lss\ out to $z = 1$ and beyond.
These same surveys, coupled with deeper x-ray surveys, should
prove profitable for the continued identification of massive clusters
with $z \simgreat 0.8$, with the corresponding implications
for structure formation models. 
The completion of several independent cluster-based peculiar velocity surveys
which all probe $\sim100-200h^{-1}$ Mpc scales but with different techniques
should, hopefully, provide a better consensus
on the convergence scale and the origin of the CMB dipole motion.
Lastly, but as important as any of the above observational efforts, the
new billion particle simulations, like those being pioneered by
the Virgo Consortium, with high spatial resolution and spanning a
large dynamic range in cosmic time will provide much more accurate and
refined model predictions. 

\section*{Acknowledgments}

I thank Tod Lauer, Michael Strauss, Istv\'an Szapudi, Michael Vogeley,
Neta Bahcall, and Harald Ebeling for the lively discussions on various aspects 
of this review. 
A special thanks to Jeff Willick for allowing me to be the first to 
publicly present his preliminary bulk flow result and to Helen Tadros for 
providing an electronic version of the APM cluster $P(k)$.


\begin{thebibliography}{999}
\bibitem{} Abadi, M. G., Lambas, D. G., , Muriel, H., 1998, ApJ, in press;
   preprint astro-ph/9806274
\bibitem{} Abell, G. O., 1958, ApJS, 3, 211 
\bibitem{} Abell, G., Corwin, H.G., Olowin, R., 1989, ApJS, 70, 1
\bibitem{} Bahcall, N., Burgett, W. S., 1986, ApJ, 300, L35
\bibitem{} Bahcall, N., West, M., 1992, ApJ, 392, 419
\bibitem{} Barnes, J., Dekel, A., Efstathiou, G., Frenk, C., 1985, ApJ,
   295, 368
\bibitem{} Boehringer, H., Collins, C., Guzzo, L., Neumann, D., 
   Schindler, S., Sch\"ucker, P., \etal, 1988, these proceedings. 
\bibitem{} Borgani, S., \etal, 1996, NewA, 1, 321 (see also astro-ph/9611100)
\bibitem{} Broadhurst, T., Ellis, R., Koo, D., Szalay, A., 1990, Nature, 
   343, 726 
\bibitem{} Broadhurst, T., \etal, 1995, in Wide Field Spectr. \& the Distant
   Universe, eds. S.J. Maddox \& A. Arag\'on-Salamanca, (Singapore: World
   Scientific), p. 178
\bibitem{} Cappi, A., Maurogordato, S., 1995, ApJ, 438, 507
\bibitem{} Carlberg, R., Morris, S., Yee, H., Ellingson, E., 1997, ApJ, 479, L19
\bibitem{} Colberg, J. M., \etal, 1998, preprint astro-ph/9808257 and these
   proceedings
\bibitem{} Couch, W.J., Ellis, R.S., Malin, D.F., \&  MacLaren, I. 1991,
   MNRAS, 249, 606
\bibitem{} Croft, R., Efstathiou, G., 1994, MNRAS, 267, 390
\bibitem{} Croft, R., Dalton, G., Efstathiou, G., Sutherland, W.,
   Maddox, S., 1997, MNRAS, 291, 305
\bibitem{dale} Dale, D., Giovanelli, R., Haynes, M., Scodeggio, M., Hardy, E.,
   Campusano, C., 1997, AJ, 114, 455
\bibitem{} Dalton, G.B., Efstathiou, G., Maddox, S.J., Sutherland, W.J.,
   1992, ApJ, 390, L1
\bibitem{} Dalton, G.B., Efstathiou, G., Maddox, S.J., Sutherland, W.J.,
   1994a, MNRAS, 269, 151 
\bibitem{} Dalton, G.B., Croft, R., Efstathiou, G., Sutherland, W., Maddox,
   S., Davis, M., 1994b, MNRAS, 271, L47 
\bibitem{} Donahue, M., Rosati, P., Dickinson, M., Postman, M., Scharf, C.,
   Mack, J., 1999, in progress
\bibitem{} Efstathiou, G., Dalton, G.B., Sutherland, W.J., Maddox, S.J., 1992,
   MNRAS, 257, 125
\bibitem{} Einasto, J., \etal, 1997, Nature, 385, 139; also astro-ph/9701018
\bibitem{} Eisenstein, D., Wayne, H., Silk, J., Szalay, A., 1998, ApJ, 494, L1
\bibitem{} Eke, V.R., Cole, S., Frenk, C., Navarro, J.F., 1996, MNRAS, 281, 703 
\bibitem{} Fan, X., Bahcall, N., Cen, R., 1997, ApJ, 490, L123
\bibitem{} Feldman, H. A., \& Watkins, R. 1994, ApJ, 430, L17
\bibitem{} Gaztanaga, E., 1994, MNRAS, 268, 913
\bibitem{} Gaztanaga, E., Croft, R., Dalton, G., 1995, MNRAS, 276, 336
\bibitem{} Geller, M., 1998, private comm.
\bibitem{} Gibbons, R., Fruchter, A., Bothun, G., 1998, private comm.
\bibitem{} Gramann, M., Bahcall, N., Cen, R., Gott, R., 1995, ApJ, 441, 449
\bibitem{} Gunn, J. E., Hoessel, J. G., Oke, J. B. 1986, ApJ, 306, 30
\bibitem{} Hamilton, A., 1993, ApJ, 417, 19 
\bibitem{} Henry, J., \etal, 1992, ApJ, 386, 408
\bibitem{} Hudson, M., \& Ebeling, H., 1997, ApJ, 479, 621
\bibitem{} Jannuzi, B., Dey, A., \etal, 1998, in progress;
   see http://www.noao.edu/noao/noaodeep/
\bibitem{} Jing, Y.P., 1997, IAU Symp. 183, 18
\bibitem{} Kaiser, N., 1984, ApJ, 284, L9 
\bibitem{} Klypin, A., Rhee, G., 1994, ApJ, 428, 399
\bibitem{} Kofman, L., Gnedin, N., Bahcall, N., 1993, ApJ, 413, 1
\bibitem{} Landy, S., Szalay, A., 1993, ApJ, 412, 64 
\bibitem{} Landy, S., Shectman, S., Lin, H., Kirshner, R., Oemler, A., 
   Tucker, D., 1996, 456, L1
\bibitem{} Lauer, T., Postman, M., 1994, ApJ, 425, 418
\bibitem{} Lauer, T., Postman, M., Strauss, M., 1999, in progress
\bibitem{} Lin, H., \etal 1996, ApJ, 471, 617
\bibitem{} Lumsden, S.L., Nichol, R., Collins, C., Guzzo, L., 1992, MNRAS, 
   258, 1 
\bibitem{} Maddox, S., Efstathiou, G., Sutherland, W., 1996, MNRAS, 283, 1227
\bibitem{} Oke, J. B., Postman, M., Lubin, L. M., 1998, AJ, 116, 549
\bibitem{} Olivier, S., Primack, J., Blumenthal, G., Dekel, A., 1993, ApJ,
   408, 17
\bibitem{} Peacock, J.A., Dodds, S.J., 1994, MNRAS, 267, 1020 
\bibitem{} Postman, M., Huchra, J., Geller, M., 1986, AJ, 91, 1267 
\bibitem{} Postman, M., Spergel, D., Sutin, B., Juszkiewicz, R., 1989, ApJ,
   346, 588
\bibitem{} Postman, M., Huchra, J., Geller, M., 1992, ApJ, 384, 404
\bibitem{} Postman, M., \etal, 1996, AJ, 111, 615
\bibitem{} Postman, M., Lubin, L. M., Oke, J. B., 1998, AJ, 116, 560
\bibitem{} Postman, M., Lauer, T., Szapudi, I., Oegerle, W., 1998b, ApJ, 
   in press; preprint astro-ph/9804141.
\bibitem{} Retzlaff, J., Borgani, S., Gottl\"ober, S., M\"uller, V., 1997,
   MNRAS, submitted; preprint astro-ph/9709044 
\bibitem{} Retzlaff, J., Borgani, S., Gottl\"ober, S., Klypin, A., M\"uller, V.,
   1998, Elsevier Preprint 
\bibitem{} Rosati, P., Della Ceca, R., Norman, C., Giacconi, R., 1998, ApJ,
   492, L21
\bibitem{} Scaramella, R., 1992, ApJ, 390, L57
\bibitem{} Scodeggio, M., Olsen, L., \etal, 1998, A\&A, submitted; preprint
   astro-ph/9807336
\bibitem{} Shectman, S., 1985, ApJS, 57, 77
\bibitem{} Sigad, Y., Eldar, A., Dekel, A., Strauss, M., Yahil, A., 1998,
   ApJ, 495, 516
\bibitem{} Soltan, A., 1988, MNRAS, 231, 309
\bibitem{} Strauss, M., Cen, R., Ostriker, J., Lauer, T., Postman, M.,
   1995, ApJ, 444, 507
\bibitem{} Sutherland, W.J., 1988, MNRAS, 234, 159 
\bibitem{} Szalay, A., Schramm, D., 1985, Nature, 314, 718
\bibitem{} Szapudi, I., Szalay, A., Bosch\'an, P., 1992, ApJ, 390, 350
\bibitem{} Tadros, H., 1998, private comm.
\bibitem{} Tadros, H., Efstathiou, G., Dalton, G., 1998, MNRAS, 296, 995
\bibitem{} Tegmark, M., Hamilton, A., Strauss, M., Vogeley, M.,
   Szalay, A., 1998, ApJ, 499, 555
\bibitem{sbf} Tonry, J., Ahjar, E., Jensen, J., Lauer, T., Postman, M.,
   Rieke, M., Thompson, R., Weyman, R., 1998, in progress (for now,
   see Lauer \etal 1997, astro-ph/9708252)
\bibitem{} Tully, B., 1987, ApJ, 323, 1
\bibitem{} Tully, B., Scaramella, R., Vettolani, G., Zamorani, G., 1992,
   ApJ, 388, 9
\bibitem{} Vikhlinin, A., \etal, 1998, ApJ, 502, 558
\bibitem{} Walter, C., Klypin, A., 1996, ApJ, 462, 13
\bibitem{wegner} Wegner, G., Colless, M., Baggley, G., Davies, R., Bertschinger,
          E., Burstein, D., McMahan, R., Saglia, R.
\bibitem{} Watkins, R., 1997, MNRAS, 292, 59
\bibitem{} Willick, J.A., 1998, ApJ, submitted; preprint astro-ph/9809160 
\bibitem{} Zwicky, F., Herzog, E., Wild, P., Karpowicz, M., Cowal, C.,
   (1961-1968), {\it Catalog of Galaxies and Clusters ...}, Vols. 1 -- 6,
   (Caltech, Pasadena) 
\end{thebibliography}
\end{document}